\documentclass[prb,a4paper,twocolumn,final]{revtex4}
\usepackage[dvips]{graphicx}

\begin{document}
\bibliographystyle{apsrev}

%\preprint{Draft V7 - cond-mat version}

\title{Radio-frequency operation of a double-island single-electron transistor}

\author{R.~Brenner}
\email[Electronic mail: ]{rolf@phys.unsw.edu.au}
\author{T.~M.~Buehler}
\author{D.~J.~Reilly}
\affiliation{Centre for Quantum Computer Technology, School of Physics, The
University of New South Wales, Sydney NSW 2052, Australia}

\date{September 2, 2004}

\begin{abstract}
We present results on a double-island single-electron transistor (DISET)
operated at radio-frequency (rf) for fast and highly sensitive detection of
charge motion in the solid state. Using an intuitive definition for the charge
sensitivity, we compare a DISET to a conventional single-electron transistor
(SET). We find that a DISET can be more sensitive than a SET for identical,
minimum device resistances in the Coulomb blockade regime. This is of
particular importance for rf operation where ideal impedance matching to
$50~\Omega$ transmission lines is only possible for a limited range of device
resistances. We report a charge sensitivity of
$5.6\times10^{-6}~e/\sqrt{\mathrm{Hz}}$ for a rf-DISET, together with a
demonstration of single-shot detection of small ($\leq0.1e$) charge signals on
$\mu\mathrm{s}$ timescales.
\end{abstract}

%\pacs{85.35.Gv, 73.23.Hk, 73.63.-b, 73.21.-b} ??

\maketitle

%%%%%%%%%%%%%%%%%%%%%%%%%%%%%%%%%%%%%%%%%%%%%%%

\section{Introduction}

With the continued miniaturization of microelectronics facing serious
limitations due to quantum effects and heat dissipation, alternative computing
paradigms are envisaged. These include solid-state quantum computers
\cite{orlov1997,shnirman1997,kane1998,makhlin1999,vrijen2000} and quantum-dot
cellular automata (QCA),\cite{lent1993} which promise a further increase in
integration and computing power beyond such limitations. In these new
architectures, readout of a computational result often requires detection of
charge motion on fast timescales -- not only to ensure fast processing speeds,
but also to perform a readout operation before the information is lost due to
interaction with the environment or relaxation. Towards this goal, the
suitability of radio-frequency single-electron transistors (rf-SETs) as fast
and sensitive charge detectors has been
demonstrated.\cite{schoelkopf1998,korotkov1999,devoret2000,aassime2001}

Recently double-island SETs (DISETs) have been proposed for readout of QCA and
qubits.\cite{macucci2001,tanamoto2003} With such applications in mind, Brenner
\textit{et al.} have shown that in addition to conventional charge sensing
(like with the single-island SET), DISETs can also be used to sense rapid
transitions between electrostatically degenerate charge
states.\cite{brenner2003a,brenner2003b} The latter, novel mode of operation
suggests that DISETs may be useful for readout of a QCA cell or a qubit in a
basis other than the charge position basis. In particular, both types of
detection can be achieved in the same device, simply by using different biasing
regimes: keeping the DISET at a high transconductance or at a conductance peak
(with zero transconductance), respectively. This versatility alone makes DISET
detectors an interesting alternative to conventional SET detectors.

In this article we report work on a radio-frequency DISET (rf-DISET), including
a measurement of its charge sensitivity and the demonstration of single-shot
measurements on $\mu\mathrm{s}$ timescales. Operation at radio-frequencies
increases the sensitivity of charge detectors as $1/f$-type noise becomes
negligible (i.e. less than the detector noise) and also facilitates the study
of processes that have short decay or relaxation times, such as qubits. We also
present an intuitive model for calculating the charge sensitivity of a DISET,
and we find that under certain conditions the charge sensitivity of a DISET is
superior to that of a SET. We consider the case where the minimum device
resistances $R_{\mathrm{d,min}}$ and source-drain biases are identical for both
devices. This is an important comparison since rf operation imposes
restrictions on the minimum device resistance needed to achieve optimum
impedance matching with a reasonable quality factor.

\section{DISET charge sensitivity}
\begin{figure}
  \center\includegraphics[width=7.62cm]{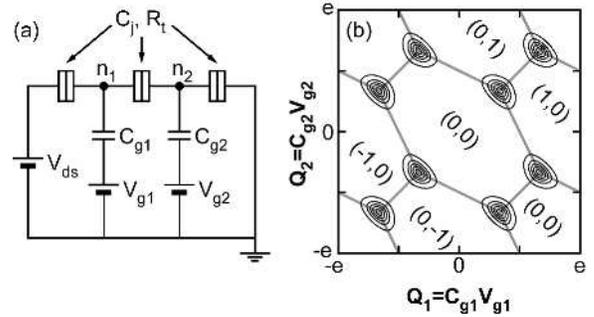}\\
  \caption{(a) Simplified circuit diagram of a DISET with identical tunnel-junctions,
biased by a source-drain voltage $V_{\mathrm{ds}}$. Gate voltages
$V_{\mathrm{g1}}$ and $V_{\mathrm{g1}}$ alter the number of excess electrons on
the two islands, $n_{1}$ and $n_{2}$. (b) The charge configurations
$(n_{1},n_{2})$ of a DISET as a function of the charges $Q_{1(2)}$ induced on
the two islands exhibit a characteristic honeycomb pattern (gray lines). At
triple-points, where three configuration domains adjoin, one can observe peaks
in the source-drain conductance (black contours; here $eV_{\mathrm{ds}}\ll
kT=E_{\mathrm{Cj}}/30=e^{2}/60C_{\mathrm{j}}$).}\label{fig:fig1ab}
\end{figure}
DISETs -- originally introduced as single-electron pumps by Pothier \textit{et
al.}\cite{pothier1992} -- consist of two small islands in series, connected by
ultra-small tunnel-junctions to each other and to source and drain leads (see
circuit diagram in Fig.~\ref{fig:fig1ab}a). Their operation is governed by
Coulomb blockade (CB) effects,\cite{grabert1992} allowing voltages applied to
capacitively coupled gates to alter the integer number of excess electrons on
island 1(2), $n_{1(2)}$, in steps of $\pm1$. Figure~\ref{fig:fig1ab}b shows
that the stable DISET charge configuration domains $(n_{1},n_{2})$ exhibit a
typical ``honeycomb'' characteristic as a function of two gate voltages,
$V_{\mathrm{g1}}$ and $V_{\mathrm{g2}}$. Current can flow through the device at
triple-points, where three charge configuration domains adjoin. Current peaks
for non-zero temperature are indicated by contours encircling the triple-points
in Fig.~\ref{fig:fig1ab}b. We refer the reader to the review by van~der~Wiel
\textit{et al.} for a more thorough discussion of the operation of
DISETs.\cite{vanderwiel2003}

Like in conventional SETs, the DISET conductance is extremely sensitive to
small induced charge signals. In the following paragraphs we give a practical,
intuitive definition for the DISET charge sensitivity and compare it to that of
a SET in order to identify potential advantages of a DISET over a SET. In
general, the charge sensitivity $\delta q$ of a detector can be defined through
its current response $\partial I/\partial Q$ to an induced charge signal in the
presence of noise, expressed by the current-noise spectral density
$S_{\mathrm{I}}$,\cite{korotkov1996}
\begin{equation}\label{eq:chsensfreq}
    \delta q=\sqrt{S_{\mathrm{I}}}
    \left(\frac{\partial I}{\partial Q}\right)^{-1}\:,
\end{equation}
or, equivalently, by the charge noise level $\Delta Q$ in a measurement
bandwidth $B$,
\begin{equation}\label{eq:chsenstime}
    \delta q=\frac{\Delta Q}{\sqrt{B}}\:.
\end{equation}
For a SET, the source-drain current is a periodic function of the induced
charge, $I_{\mathrm{SET}}(Q)=I_{\mathrm{SET}}(Q+ne)$, where $n$ is the number
of excess electrons on the SET island. A DISET on the other hand has two
islands, which sense an effective charge $Q_{1}$ and $Q_{2}$, respectively. The
DISET current is thus a periodic function of the induced charges on both
islands,
$I_{\mathrm{DISET}}(Q_{1},Q_{2})=I_{\mathrm{DISET}}(Q_{1}+n_{1}e,Q_{2}+n_{2}e)$.
In an idealized case, the effective induced charge on one island is
proportional to the other, $Q=Q_{1}=\gamma Q_{2}$. Two special cases are
considered here: $\gamma=+1$ (e.g. single gate or chargeable object coupled
equally strongly to both islands) and $\gamma=-1$ (e.g. polarisable object
parallel to the DISET, such as the double-dot in Fig.~\ref{fig:fig3} and
Refs.~\onlinecite{brenner2003a,brenner2003b}). The current is then a function
of the induced charge on only one of the islands for both values of $\gamma$,
$I_{\mathrm{DISET}}^{\pm}(Q,\pm Q)=I_{\mathrm{DISET}}^{\pm}(Q)$, and $\partial
I_{\mathrm{DISET}}^{\pm}/\partial Q$ can be considered the response to a
relevant induced charge signal $Q$.\cite{othercharge}
\begin{figure}
  \center\includegraphics[width=8.47cm]{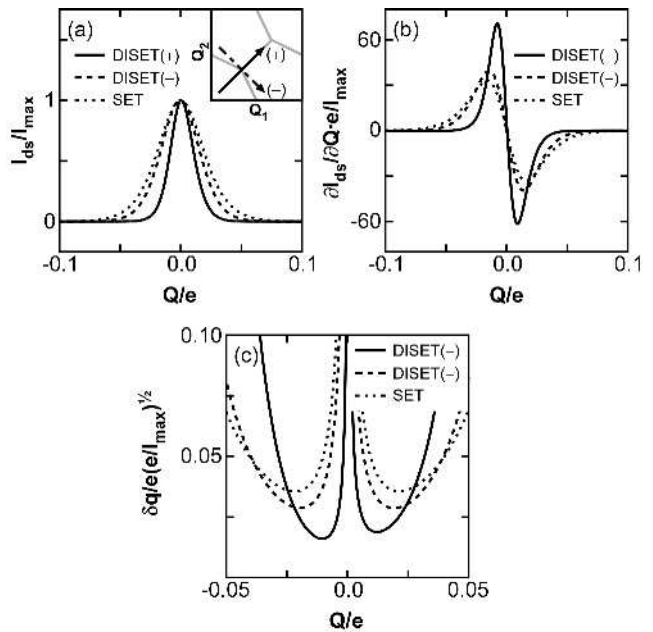}\\
  \caption{(a) Calculated dc current peaks for a SET and a DISET for $\gamma=+1$ ($+$) and
$\gamma=-1$ ($-$) and normalized to the maximum peak current
$I_{\mathrm{max}}$. The DISET traces are cross-sections across a triple-point
of the honeycomb diagram in two directions (see inset). (b) Differentiation of
(a) yields a maximum differential current (normalized to $I_{\mathrm{max}}/e$)
for the DISET with $\gamma=+1$. (c) Using (a) and (b), the shot-noise limited
charge sensitivities are obtained, normalized to $e/\sqrt{I_{\mathrm{max}}/e}$
and neglecting shot noise suppression effects. For identical
$I_{\mathrm{max}}$, the DISET has a higher charge sensitivity than the SET by a
factor of up to 2.23.}\label{fig:fig2abc}
\end{figure}

Based on this definition of the relevant induced charge, we calculate the
(normal-state) dc source-drain current $I_{\mathrm{ds}}$ and differential
current $\partial I_{\mathrm{ds}}/\partial Q$ of a DISET and compare them to
those of a SET. The dc source-drain currents are obtained from the stationary
solutions of the respective master equations in the two-state (SET) or
three-state (DISET) approximation.\cite{mastereq} Here, the following
assumptions are made:
\begin{itemize}
    \item All tunnel junctions capacitances $C_{\mathrm{j}}$ are identical;
    \item The minimum device resistances $R_{\mathrm{d,min}}=
    n^{2}R_{\mathrm{t}}$ for the SET and the DISET in the CB regime are identical
    ($R_{\mathrm{t}}$ is the resistance of a single junction and $n$ is the number
    of junctions in the respective device);
    \item The absolute induced charges $Q$ are identical for all islands;
    \item The temperature is low compared to the
    charging energy of a single junction: $kT\ll E_{\mathrm{C}}=e^{2}/2C_{\mathrm{j}}$
    (here $T=E_{\mathrm{C}}/50k$);
    \item The source-drain bias is small -- i.e. $V_{\mathrm{ds}}\ll kT/e$ (linear
    response) -- and identical for both devices; and
    \item Cotunneling is disregarded.
\end{itemize}
Figure~\ref{fig:fig2abc}a shows the respective centered dc current peaks for a
SET and a DISET for $\gamma=+1$ and $\gamma=-1$ as a function of $Q$. The
currents are normalized to the maximum peak current
$I_{\mathrm{max}}=V_{\mathrm{ds}}/R_{\mathrm{d,min}}$. In the case of a DISET,
the induced charges on the islands are $Q=Q_{1}=\pm Q_{2}$; i.e. a triple-point
in the honeycomb diagram of the DISET is traversed either from the top-left to
the bottom-right ($-$) or from the bottom-left to the top-right ($+$), as
indicated in the inset. The differential currents (normalized to
$I_{\mathrm{max}}/e$) are shown in Fig.~\ref{fig:fig2abc}b, indicating that the
maximum value is achieved for a DISET with $\gamma=+1$. Note that the DISET
peak for $\gamma=+1$ is asymmetric and has a higher transconductance on one
side of the peak than on the other. This needs to be taken into consideration
measurements that require maximum charge sensitivity.

Ultimately, the noise in single-electron devices is limited by shot noise with
the noise-current spectral density
\begin{equation}\label{eq:schottky}
    S_{\mathrm{I}}=2\eta eI_{\mathrm{ds}}\:,
\end{equation}
where $\eta$ is the Fano factor accounting for shot noise suppression or
enhancement effects. The charge sensitivities as a function of $Q$ where
calculated for $\eta=1$. Figure~\ref{fig:fig2abc}c shows that the charge
sensitivity of a DISET is better than that of a SET by a factor of up to $2.23$
for identical maximum peak currents $I_{\mathrm{max}}$ and for the assumptions
made above. This is of particular interest in light of the restrictions for
$R_{\mathrm{d,min}}$ associated with rf operation. The intuitive picture
presented here was used as a basis to determine the charge sensitivity of a
rf-DISET for $\gamma=+1$ experimentally.

There are two other effects, which are not included in the model presented here
and which may also lead to an enhanced sensitivity for the DISET in comparison
to single-island devices. Firstly, the different Fano factors for DISETs and
SETs may lead to a further reduction of shot noise in the DISET compared to the
SET. The shot noise level in DISETs can be suppressed by a factor of up to
$\eta=1/3$,\cite{eto1996,korotkov2000,gattobigio2002} compared to $\eta=1/2$
for SETs.\cite{hanke1994,birk1995} Secondly, additional shot noise due to
cotunneling is also suppressed in DISETs (compared to SETs) due to the
additional tunnel junction.\cite{averin1989,averin1990} Both of the above
effects are strongly dependent on the source-drain bias, the induced charge and
the junction parameters.

\section{rf-DISET setup and operation}
\begin{figure}
  \center\includegraphics[width=8.15cm]{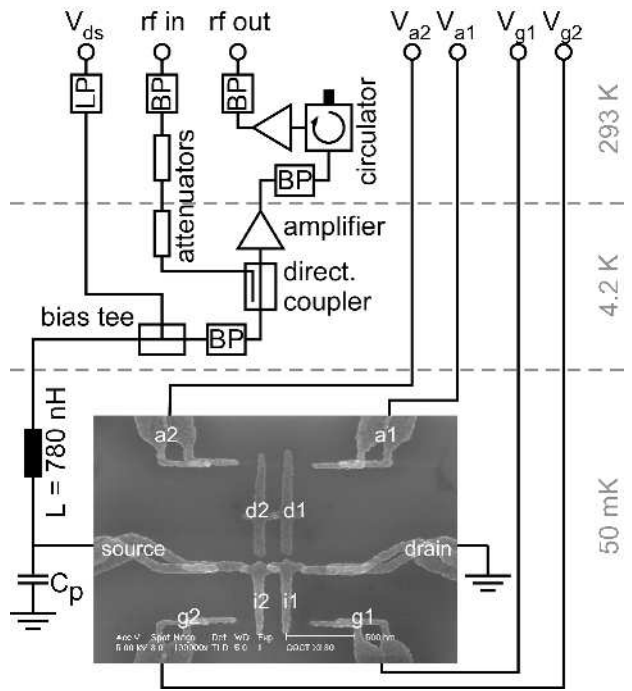}\\
  \caption{Schematic of the electronic setup and scanning electron micrograph of a
rf-DISET. The rf carrier signal is band-pass (BP) filtered and attenuated at
various stages, and the dc source-drain bias is low-pass (LP) filtered before
being added through a bias-tee. The signal reflected off the LCR tank circuit
is again BP filtered and amplified by a cryogenic and a room-temperature rf
amplifier. Voltages applied to gates g1, g2, a1 and a2 capacitively couple to
the islands i1 and i2 and can alter the DISET charge configurations. The
double-dot (comprising dots d1 and d2) was not used in the experiments reported
here.}\label{fig:fig3}
\end{figure}

Motivated by the possibility of increased sensitivity, we now turn to discuss
our experiment on the rf-DISET. The experimental setup employed for this work
is schematically shown in Fig.~\ref{fig:fig3}, which also shows a scanning
electron micrograph of the DISET device (the double-dot adjacent to the DISET
was not used in these experiments). The device was fabricated using standard
shadow evaporation of aluminum with \textit{in-situ}
oxidation.\cite{fulton1987}

Fast operation is made possible by including the DISET in an impedance-matching
LCR tank circuit -- a technique pioneered by Schoelkopf \textit{et al.} for
conventional SETs made of aluminum.\cite{schoelkopf1998} The rf carrier signal
and a dc source-drain bias were coupled into an impedance matching LCR tank
circuit,\cite{pozar1998} consisting of commercially available chip inductors
(impedance $L$), a parasitic capacitance to ground
($C_{\mathrm{p}}\approx0.23~\mathrm{pF}$) and the total DISET resistance
$R_{\mathrm{d}}$. Using an impedance of $L=780~\mathrm{nH}$, the resonance
frequency of the tank circuit was set to
$f_{0}=1/(2\pi\sqrt{C_{\mathrm{p}}L})\approx375.8~\mathrm{MHz}$ within the
operating bandwidth of the rf amplifiers. The reflection coefficient
$\Gamma^{2}$ for the rf signal reflected from the tank circuit is a function of
$R_{\mathrm{d}}$ (as are the quality factor and the bandwidth of the tank
circuit). In the case of perfect matching, $\Gamma^{2}=0$ when the DISET is at
a conductance maximum and $\Gamma^{2}=1$ when it is blockaded. The optimization
of the matching and the performance of the detector system as a whole imposes
limitations on the choices for the minimum device resistance
$R_{\mathrm{d,min}}$ (here: perfect matching for
$R_{\mathrm{d,min}}=67.8~\mathrm{k}\Omega$). The device presented in this paper
had a minimum device resistance of
$R_{\mathrm{d,min}}\approx189.8~\mathrm{k}\Omega$ in the CB regime, which
resulted in a slightly mismatched tank-circuit.

The reflected rf signal was amplified by a cryogenic, ultra-low-noise rf
amplifier \cite{berkshire2004} and a low-noise rf amplifier at
room-temperature. A circulator prevented noise at the front-end of the
room-temperature amplifier from propagating to the cryogenic amplifier. For
time-domain measurements, the signal was demodulated and fed into a
multichannel oscilloscope, while for frequency-domain measurements, the
reflected signal was fed into a spectrum analyzer. The gates were modulated on
fast timescales using a synthesized function generator and/or on slow
timescales using digital-to-analog converters. A magnetic field of 0.5~T was
applied to suppress superconductivity in the aluminum.

\begin{figure}
  \center\includegraphics[width=8cm]{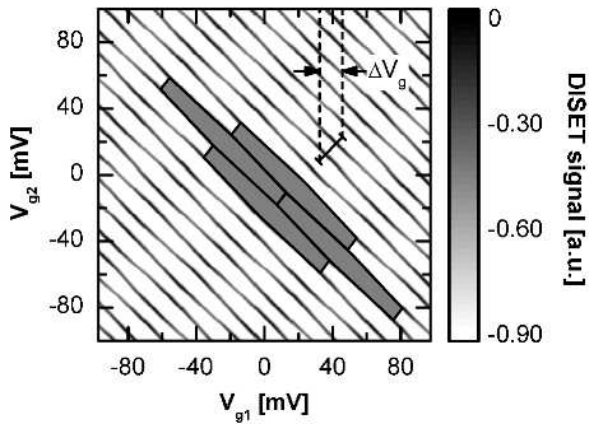}\\
  \caption{Demodulated, reflected rf signal exhibiting the characteristic honeycomb
structure as a function of gate voltages $V_{\mathrm{g1}}$ and
$V_{\mathrm{g2}}$. Gate g1 was modulated with a triangular wave at 962~Hz,
while gate g2 was stepped. Each trace was averaged 256 times. Four hexagonal
cells are highlighted by gray shading, and white regions correspond to CB,
while conductance maxima are shaded in black.}\label{fig:fig4}
\end{figure}

By applying a triangular modulation signal to gate g1
($f_{\mathrm{g}}=962~\mathrm{Hz}$) and stepping $V_{\mathrm{g2}}$ on a 1-s
timescale, the characteristic honeycomb diagram of the DISET was obtained from
the demodulated, reflected rf signal (Fig.~\ref{fig:fig4}). The observed
non-zero conductance (gray) along the edges of the honeycombs is likely due to
cotunneling. This is consistent with relatively low junction resistances of
order $15\ldots25~\mathrm{k}\Omega$, estimated from the room-temperature
resistance of $\sim70~\mathrm{k}\Omega$.

\section{Experimental results}
To determine the charge sensitivity of the rf-DISET, its response (in terms of
the reflected power) to a g-gate signal with amplitudes
$V_{\mathrm{g}}=V_{\mathrm{g1}}=V_{\mathrm{g2}}$ was measured in both the
frequency domain and the time domain. The period of CB oscillations for this
case, $\Delta V_{\mathrm{g}}=\Delta V_{\mathrm{g1}}=\Delta V_{\mathrm{g2}}$
(indicated in Fig.~\ref{fig:fig4}), corresponded to the addition of one
electron to each island. Hence, the effective induced charge was given by
$Q=eV_{\mathrm{g}}/\Delta V_{\mathrm{g}}$. Constant offset voltages were
applied to the a-gates to bias the DISET to a point of maximum
transconductance.

\begin{figure}
  \center\includegraphics[width=8cm]{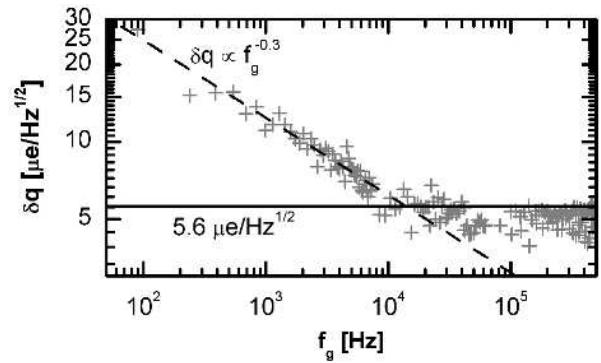}\\
  \caption{As a function of the modulation frequency, the charge sensitivity shows the
effect of flicker noise at low frequencies (with a $f_{\mathrm{g}}^{-0.6}$
dependence of the noise power). The charge sensitivity was obtained from the
SNR of amplitude modulation sidebands, and was optimum for
$f_{\mathrm{g}}>50~\mathrm{kHz}$ with $\delta
q\approx5.6\times10^{-6}~e/\sqrt{\mathrm{Hz}}$.}\label{fig:fig5}
\end{figure}

For measurement in the frequency-domain, an equivalent $Q=0.05e$ (rms) signal
was applied to the g-gates at various modulation frequencies $f_{\mathrm{g}}$,
and using a carrier signal at the resonance frequency $f_{0}$ of the tank
circuit. A spectrum analyzer was used to measure the signal-to-noise ratio
(SNR) of the amplitude modulation side-bands. Using
\begin{equation}\label{eq:chargesensmeas}
    \delta q=\frac{Q}{\sqrt{B}\times10^{\mathrm{SNR}/20}}\:,
\end{equation}
the charge sensitivity was obtained as a function of $f_{\mathrm{g}}$
(Fig.~\ref{fig:fig5}). The data shown contains a number of individual data
sets, which were recorded at different resolution bandwidths $B$ of the
spectrum analyzer. Due to uncertainties in the noise level, the worst obtained
SNR was considered the most meaningful figure when determining the actual
charge sensitivity. For modulation frequencies below $\sim30~\mathrm{kHz}$, the
charge sensitivity was reduced due to $1/f$-type noise -- here, with a
$f_{\mathrm{g}}^{-0.6}$ dependence of the noise power. Above approximately
50~kHz, amplifier noise dominated, resulting in an estimated charge sensitivity
of $\delta q\approx5.6\times10^{-6}~e/\sqrt{\mathrm{Hz}}$.

This initial result is comparable to the best charge sensitivities obtained in
rf-SET setups to date: The best such sensitivity of $\delta
q\approx3.2\times10^{-6}~e/\sqrt{\mathrm{Hz}}$ was achieved by Aassime
\textit{et al.} in a rf-SET in the superconducting
state.\cite{aassime2001,aassime2002} Buehler \textit{et al.} achieved charge
sensitivities of $4.4\times10^{-6}~e/\sqrt{\mathrm{Hz}}$ and
$7.5\times10^{-6}~e/\sqrt{\mathrm{Hz}}$ in two simultaneously operated rf-SETs
in a twin-rf-SET experiment.\cite{buehler2003b,buehler2003d,buehler2003e} The
DISET charge sensitivity reported here is all the more encouraging since the
junction parameters were not ideal for optimum device performance. Better
impedance matching of the tank circuit also promises further improvement of the
DISET charge sensitivity.

\begin{figure}
  \center\includegraphics[width=7cm]{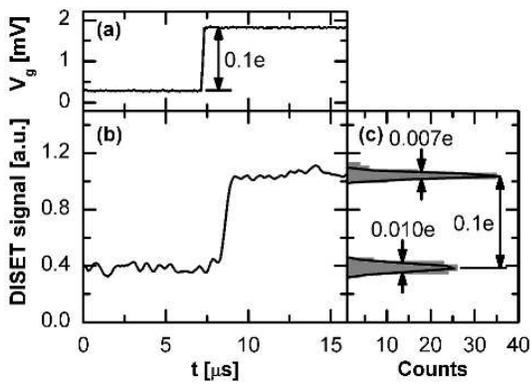}\\
  \caption{(a) Square-wave g-gate signal with $0.1e$ equivalent amplitude and (b) response
of the DISET. (c) From the widths of the peaks in the histogram obtained from
the data in (b), the time-domain charge sensitivity was estimated to $\delta
q\approx1.7\times10^{-5}~e/\sqrt{\mathrm{Hz}}$.}\label{fig:fig6abc}
\end{figure}

In order to demonstrate charge detection on fast timescales, a square wave
signal with 50~kHz modulation frequency and $Q=0.1e$ equivalent amplitude was
applied to the g-gates (Fig.~\ref{fig:fig6abc}a). The demodulated, reflected
signal was amplified and low-pass filtered (1~MHz) before being detected by an
oscilloscope. Figure~\ref{fig:fig6abc}b shows the response of the DISET in a
single-shot measurement for a measurement time of
$t_{\mathrm{meas}}=16~\mu\mathrm{s}$. From the average half-width of the
Gaussian distributions fitted to the histogram of the data
(Fig.~\ref{fig:fig6abc}c), the charge noise was estimated to $\Delta
Q\approx8.5\times10^{-3}e$. With the measurement bandwidth
$B=1/t_{\mathrm{meas}}=62.5~\mathrm{kHz}$ and Eq.~(\ref{eq:chsenstime}), an
estimate for the time-domain charge sensitivity of $\delta
q\approx1.7\times10^{-5}~e/\sqrt{\mathrm{Hz}}$ was obtained.

Longer measurement times are required in order to detect smaller charge signals
with a minimum error rate. In the presence of $1/f$ type noise, however, the
Gaussian distributions increasingly overlap with measurement time, which
increases the error probability of a single-shot measurement and in turn
imposes limitations on the smallest detectable charge signals for a given,
tolerable error probability \cite{buehler2003b,buehler2003c}. The discrepancy
to the frequency-domain charge sensitivity can be attributed to additional
noise in the demodulation circuit and the oscilloscope used in time-domain
measurements. These noise contributions have to be minimized in order to obtain
a time-domain sensitivity similar to the frequency-domain sensitivity. This was
achieved to a good degree by Bladh \textit{et al.} who reported a time-domain
sensitivity of approximately $6.3\times10^{-6}~e/\sqrt{\mathrm{Hz}}$, compared
to a frequency-domain sensitivity of
$5.5\times10^{-6}~e/\sqrt{\mathrm{Hz}}$.\cite{bladh2003} We also note that the
frequency-domain charge sensitivity of the present rf-DISET was not limited by
shot noise but rather by amplifier noise and other noise sources in the rf
network.

In order to quantify the system noise, the noise temperature was measured
according to the method described in Ref.~\onlinecite{aassime2001} and assuming
a Fano factor of $\eta=1/3$. We obtain a noise temperature of
$T_{\mathrm{n}}\approx2.1~\mathrm{K}$, which is lower than previously reported
noise temperatures of
$\geq3.5~\mathrm{K}$.\cite{aassime2001,buehler2003b,roschier2004} The
individual contributions of components to the noise in the rf network were not
explicitly studied in this work. However, the largest contribution stems from
the cryogenic amplifier (according to its specifications approximately 1.6~K at
20~K and possibly less at 4.2~K). The remaining contributions are attributed to
the imperfectly matched impedance of the tank circuit to the rf network, the
room-temperature rf amplifier and losses and reflections elsewhere in the
network.

\section{Conclusion}
Motivated by a possible enhancement in charge sensitivity in comparison to
conventional SETs, we have presented experimental results on a novel rf-DISET
detector. In certain circumstances we find that DISETs can improve the charge
sensitivity by a factor of 2.23 without increasing the minimum device
resistance required for optimum impedance matching. Our experimental results
indicate a charge sensitivity of $5.6\times10^{-6}~e/\sqrt{\mathrm{Hz}}$, and a
single-shot time-domain sensitivity of $1.7\times10^{-5}~e/\sqrt{\mathrm{Hz}}$.
The noise temp of our entire system was measured to be 2.1~K. The capability of
DISETs to detect both well-defined and electrostatically degenerate charge
configurations, may make DISETs an interesting alternative to conventional SETs
for readout of QCA-type devices and qubits in solid-state quantum computer
architectures.

\section*{Acknowledgements}
The authors would like to thank R.~P. Starrett and D. Barber for technical
assistance and A.~D. Greentree, A.~R. Hamilton and  A.~N. Korotkov for helpful
discussions. This work was supported by the Australian Research Council, the
Australian government and by the US National Security Agency (NSA), Advanced
Research and Development Activity (ARDA) and the Army Research Office (ARO)
under contract number DAAD19-01-1-0653.

%%%%%%%%%%%%%%

%%%%%%%%%%%%%%


\begin{thebibliography}{37}
\bibitem[{mas()}]{mastereq}
\bibinfo{note}{For a SET, only transitions between SET charge states $(n)$ and
  $(n+1)$ were considered. For a DISET, transitions between charge states
  $(n_{1}\pm1,n_{2})$, $(n_{1},n_{2})$ and $(n_{1},n_{2}\pm1)$ were considered,
  where the plus and minus cases were treated separately and correspond to two
  neighbouring triple-points.}
\end{thebibliography}
\end{document}